\documentclass[%
 reprint,
 superscriptaddress,
 amsmath,amssymb,
 aps,prb
]{revtex4-1}

\usepackage[utf8x]{inputenc}
\usepackage{graphicx}% Include figure files
\usepackage{dcolumn}% Align table columns on decimal point
\usepackage{bm}% bold math
\usepackage{hyperref}% add hypertext capabilities
\usepackage{textgreek}
\usepackage{braket}
\usepackage{xr}

\begin{document}

\title{Phonon-assisted Exciton Dissociation in Transition Metal Dichalcogenides}

\author{Raül Perea-Causín}
\author{Samuel Brem}
\author{Ermin Malic}
\affiliation{Department of Physics, Chalmers University of Technology, 412 96 Gothenburg, Sweden}

\begin{abstract}
Monolayers of transition metal dichalcogenides (TMDs) have been established in the last years as promising materials for novel optoelectronic devices. However, the performance of such devices is often limited by the dissociation of tightly bound excitons into free electrons and holes. While previous studies have investigated tunneling at large electric fields, we focus in this work on phonon-assisted exciton dissociation that is expected to be the dominant mechanism at small fields. We present a microscopic model based on the density matrix formalism providing access to time- and momentum-resolved exciton dynamics including phonon-assisted dissociation. We track the pathway of excitons from optical excitation via thermalization to dissociation, identifying the main transitions and dissociation channels.
Furthermore, we find intrinsic limits for the quantum efficiency and response time of a TMD-based photodetector and investigate their tunability with externally accessible knobs, such as excitation energy, substrate screening, temperature and strain. Our work provides microscopic insights in fundamental mechanisms behind exciton dissociation and can serve as a guide for the optimization of TMD-based optoelectronic devices.\\[4pt]
\end{abstract}

\maketitle

%\tableofcontents

%%%%%%%%%%%%%%%%%%%%%% INTRODUCTION %%%%%%%%%%%%%%%%%%%%%%%%%
The extensive research on two-dimensional materials during the last decade has put monolayers of transition metal dichalcogenides (TMDs) in the spotlight for next-generation optoelectronic applications\,\cite{mak2016photonics,pospischil2016optoelectronic,wang2018colloquium,mueller2018exciton}. The strong light-matter interaction and the ultrafast non-equilibrium dynamics\,\cite{mak2010atomically,splendiani2010emerging,chernikov2014exciton,berghauser2014analytical,he2014tightly} makes them excellent candidates for active materials in photodetectors and solar cells.
The understanding of their properties has significantly advanced in the last years with  experimental and theoretical studies shining light on the optical response\,\cite{moody2015intrinsic,selig2016excitonic,steinhoff2015efficient,brem2019intrinsic,raja2019dielectric}, exciton relaxation dynamics\,\cite{selig2018dark,brem2018exciton,miller2017long,rosati2020temporal} and exciton propagation\,\cite{cadiz2018exciton,perea2019exciton,leon2019hot,rosati2020negative,zipfel2020exciton} in different TMD materials. While these are  important properties for the operation of a photodetector, exciton dissociation plays a central role as the bridge between optical excitation and photocurrent generation. 

\begin{figure}[t!]
 \centering
 \includegraphics[width=\linewidth]{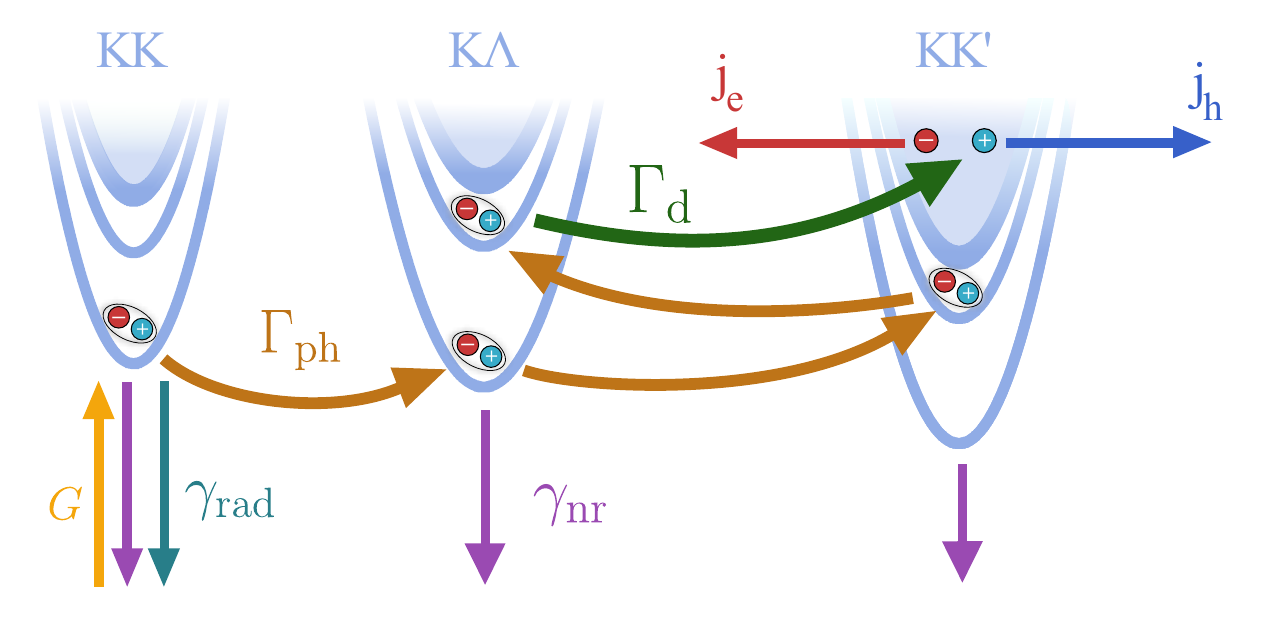}
 \caption{\textbf{Exciton dissociation}. Schematic representation of exciton thermalization and dissociation in WSe\textsubscript{2}. Excitons are generated by optical excitation in the KK valley (represented by $G$ in the figure). Scattering with phonons leads to a thermalization of excitons and a redistribution across all availbale states ($\Gamma_{\text{ph}}$). Excitons close to the continuum can dissociate into free electrons and holes ($\Gamma_{\text{d}}$) preserving the thermal equilibrium between bound excitons and unbound carriers. Finally, free carriers are dragged away by an electric field and produce a photocurrent ($j_{\text{e(h)}}$). Excitons in the light cone at the KK valley can recombine radiatively ($\gamma_{\text{rad}}$), while excitons with non-zero momentum  recombine non-radiatively ($\gamma_{\text{nr}}$).}
 \label{fig:Fig1}
\end{figure}

During the last years the characteristics of TMD-based optoelectronic devices have been investigated, reporting promising features, such as fast photoresponse\,\cite{massicotte2016picosecond,wang2015ultrafast}, large responsivities\,\cite{lopez2013ultrasensitive,furchi2014mechanisms,pospischil2014solar,furchi2018device} and high tunability\,\cite{baugher2014optoelectronic,zhang2013high}. To reach optimal operation of these devices,  exciton dissociation mechanisms and their tunability need to be understood. Previous studies have investigated exciton dissociation via tunneling to the continuum in presence of strong electric fields\,\cite{haastrup2016stark,kamban2019field}, showing that field-induced dissociation dominates the response at large fields needed to dissociate excitons with large binding energies\,\cite{massicotte2018dissociation}. However, the dissociation of excitons via scattering with phonons, which is expected to play a major role at low electric fields, has been overlooked so far---despite its potential implications on the fundamental efficiency limits of TMD-based photodetectors and solar cells.
After an optical excitation, the generated excitons scatter with phonons to reach a thermal equilibrium leading to a finite population of unbound electrons and holes as described by the Saha equation\,\cite{steinhoff2017exciton}. Under the presence of an electric field, these free carriers are dragged away and excitons dissociate to preserve the thermal equilibrium (cf. Fig\,\ref{fig:Fig1}). Exciton dissociation hence poses an upper fundamental limit on the efficiency of TMD-based photodetectors and solar cells.

The aim of this work is to provide microscopic insights in fundamental processes governing  exciton dissociation in  TMD materials. 
Based on a fully quantum mechanical approach, we  resolve the dissociation dynamics of excitons in energy, momentum and time.
We take into account the Rydberg-like series of bright and dark excitonic states giving rise to a multi-excitonic landscape, which plays an essential role for the dissociation process.
We track the pathway of excitons from optical excitation via thermalization to dissociation into free carriers and reveal the underlying microscopic channels (cf. Fig\,\ref{fig:Fig1}). In particular, we find that intervalley scattering dominates the dissociation dynamics due to its strong exciton-phonon coupling, scattering excitons to higher energies and dissociating them into free electrons and holes.
Furthermore, we investigate the tunability of the dissociation efficiency and response time through externally accessible knobs and find the best conditions for optimal operation of a TMD-based photodetector. We find a dissociation-limited response time for WSe$_2$ monolayers that is in good agreement to reported experimental values\,\cite{massicotte2018dissociation}, supporting the accuracy of our model and the crucial role of exciton-phonon scattering at weak fields.\\

%%%%%%%%%%%%%%% THEORY %%%%%%%%%%%%%%%%%%%%%%%%%%%%%%%%%%%%%%%%%%%%%%%%%%%%%%%%%%%%%%%%%%%%%%%%%
\noindent\textbf{Microscopic model.}
The photocurrent created through dissociation of excitons is the key quantity in TMD-based optoelectronic devices, such as photodetectors and solar cells. In excitonic basis, the current can be expressed as $\braket{\bm{j}} \propto \sum_{\lambda \bm{k}} \bm{k} f^{\lambda}_{\bm{k}} \propto \sum_{\nu \bm{k}} \bm{k} |\phi^{\nu}_{\bm{k}}|^2 n_{\nu}$, where $f^{\lambda}_{\bm{k}}$ is the single-particle occupation, $\phi^{\nu}_{\bm{k}}$ the exciton wavefunction, and $n_{\nu}$ the exciton density in the state $\nu$.
Since the squared wavefunction $|\phi^{\nu}_{\bm{k}}|^2$ for bound states is an even function and $\bm{k}$ is odd, the momentum integration yields zero current. This is not the case for unbound electron-hole pairs, and hence the photocurrent is generated only by free carriers, cf. the supplementary material.
When TMD-based devices are optically excited, excitons are formed and scatter with phonons to reach a thermal distribution with a finite population of unbound carriers. Under the presence of an electric field, free carriers are dragged away generating a current and excitons close to the continuum continuously dissociate preserving the thermal equilibrium.
The dissociation process can be slow compared to transport and therefore poses a fundamental limit to the current generation and the response time of photodetectors. In the following, we present a quantum-mechanical model to describe the dynamics of excitons including optical exciation, intra- and intervalley thermalization and, in particular, exciton dissociation.

Based on the density matrix formalism in second quantization\,\cite{haug2009quantum,kira2006many,kira2011semiconductor,malic2013graphene}, we make use of the many-particle Hamilton operator and the Heisenberg's equation of motion to describe the dynamics of incoherent excitons $N^{\nu}_{\bm{Q}} = \braket{X^{\nu \dagger}_{\bm{Q}} X^{\nu\phantom{\dagger}}_{\bm{Q}}}$.
Here, we have introduced the exciton operator $X^{\nu (\dagger)}_{\bm{Q}}$ accounting for the annihilation (creation) of an exciton in the state $\nu$ with momentum $\bm{Q}$.
The Hamilton operator including electron-electron, electron-phonon, and electron-photon interactions is  transformed into an excitonic basis resulting in an excitonic Hamiltonian\,\cite{katsch2018theory}.
We focus on low-density conditions, where exciton-exciton interaction is negligible\,\cite{moody2015intrinsic,perea2019exciton}.
The central part of our work lies in the exciton-phonon interaction that drives exciton thermalization and dissociation. In general, the exciton-phonon Hamiltonian describing scattering from $\ket{\mu,\bm{Q}}$ to $\ket{\nu,\bm{Q}+\bm{q}}$ has the form\,\cite{brem2019intrinsic}
\begin{equation}
 H_{\text{x-ph}} = \sum_{j\nu\mu\bm{Qq}} G^{\mu\nu}_{j\bm{q}} X^{\nu\dagger}_{\bm{Q}+\bm{q}} X^{\mu\phantom{\dagger}}_{\bm{Q}} \left(b^{j\phantom{\dagger}}_{\bm{q}}+b^{j\dagger}_{-\bm{q}}\right),
\end{equation}
where we have introduced the annihilation (creation) operator $b^{j(\dagger)}_{\bm{q}}$ for phonons with the mode $j$ and the momentum $\bm{q}$. The exciton-phonon matrix element $G^{\mu\nu}_{j\bm{q}} = \mathcal{F}^{\mu\nu}_{\alpha_h\bm{q}} g^c_{j\bm{q}} - \mathcal{F}^{\mu\nu}_{-\alpha_e\bm{q}} g^v_{j\bm{q}}$ contains the excitonic form factor $\mathcal{F}^{\mu\nu}_{\bm{q}} = \sum_{\bm{k}} \phi^{\mu *}_{\bm{k}} \phi^{\nu}_{\bm{k}+\bm{q}}$, the electron-phonon coupling strength $g^{\lambda}_{j\bm{q}}$ and the factor $\alpha_{\lambda}=m_{\lambda}/(m_h+m_e)$. 
The excitonic wavefunctions $\phi^{\nu}_{\bm{k}}$ are obtained by solving the Wannier equation with a thin-film Coulomb potential\,\cite{brem2019intrinsic,rytova1967,keldysh}.
We describe excitonic states with positive binding energy (free states) by orthogonal plane waves (OPW)\,\cite{bockelmann1992electron,schneider2001many}. Free states form a continuum in which the quantum number $\nu$ becomes a continuous momentum $\bm{p}$ with the energy $E^{\bm{p}}_{\bm{Q}} = E_{\text{gap}}+\frac{\hbar\bm{Q}^2}{2M}+\frac{\hbar\bm{p}^2}{2m_r}$ (in an effective mass approximation with $m_r$ being the reduced mass and $M$ the total mass).
The OPWs describing the states in the continuum can be written as $\psi^{\bm{p}}_{\bm{k}} = \delta_{\bm{pk}} - \sum_{\nu}^{N_{\text{b}}} \phi^{\nu*}_{\bm{p}} \phi^{\nu}_{\bm{k}}$, where $N_{\text{b}}$ is the number of bound states. While the first term corresponds to the plane wave description of continuum states, the second term accounts for the orthogonalization of these states with respect to the bound states in order to avoid unphysical wavefunction overlaps, cf. the supplementary material.

\begin{figure*}[t!]
 \centering
 \includegraphics[width=\linewidth]{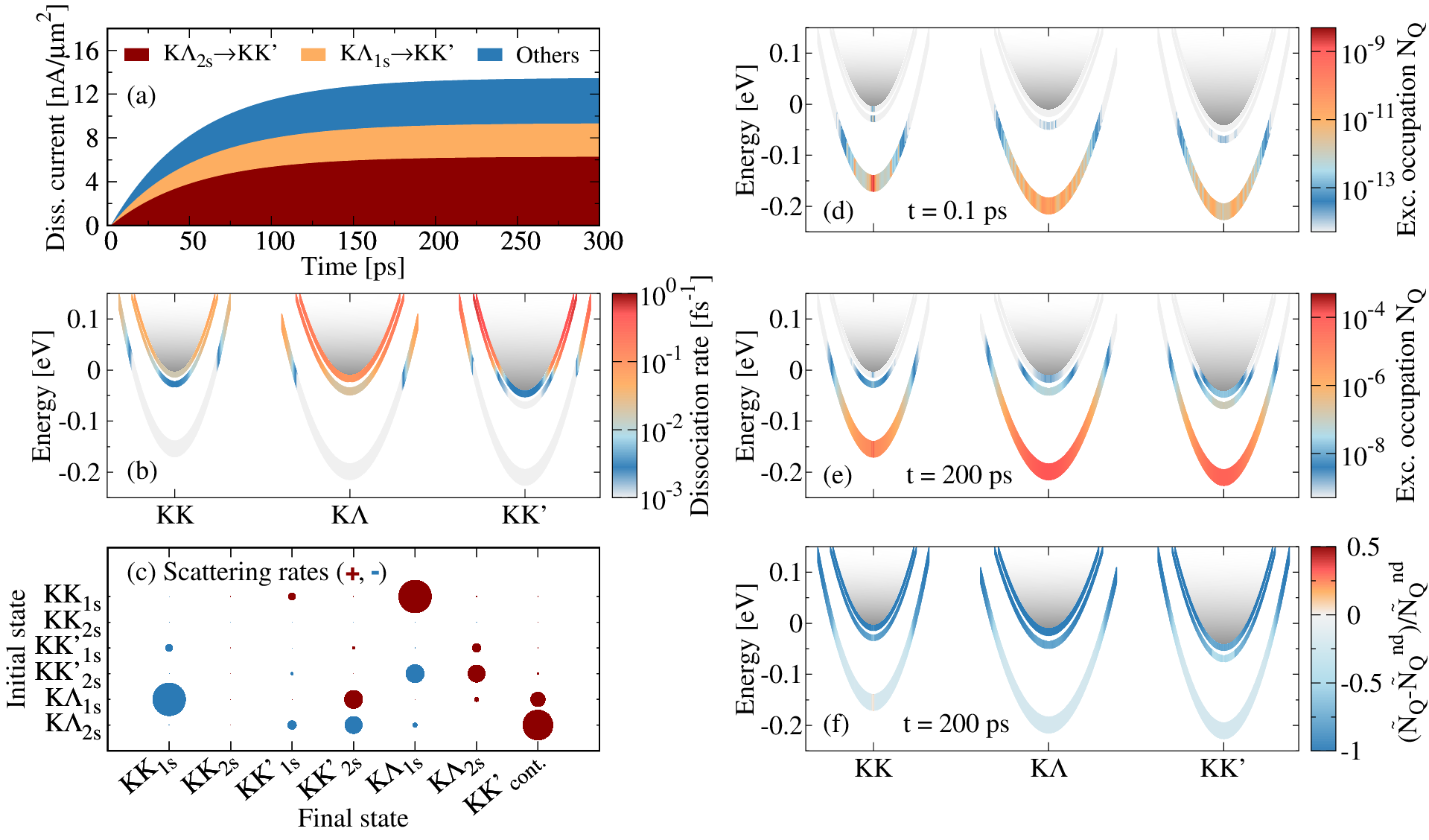}
 \caption{\textbf{Dissociation dynamics}. (a) Dissociation current density $j_{\text{d}}$ in hBN-encapsulated WSe\textsubscript{2} as a function of time with disentangled contributions from the most important channels. A typical non-radiative decay time $\tau_{\text{nr}} = 100\,\text{ps}$ has been considered.
 (b) Dissociation rates $\Gamma^{\text{diss}}_{\nu\bm{Q}}$ as a function of momentum and energy. The grey gradient at high energies illustrates free states and the energy axis is shifted, so that the KK continuum is located at 0 eV. 
 (c) Transition rate $\Gamma_{\nu\mu}$ from specific excitonic states into other bound states or the continuum. The size of the circles is proportional to the corresponding rate. Positive and negative values are represented by red and blue, respectively.
Exciton occupation as a function of momentum and energy at (d) $t = 0.1\,\text{ps}$  and (e) $t = 200\,\text{ps}$ after optical excitation.
 (f) Relative difference between the exciton occupation with ($N_{\bm{Q}}$) and without ($N^{\text{nd}}_{\bm{Q}}$) considering dissociation. 
}
 \label{fig:Fig2}
\end{figure*}

Using the Heisenberg's equation of motion we obtain an expression for momentum- and time-resolved exciton occupation:
\begin{align}
 \dot{N}^{\nu}_{\bm{Q}}(t) & = \sum_{\mu} \Gamma^{\mu\nu}_{\bm{0Q}} |p^{\mu}(t)|^2 - 2 \delta_{\bm{Q0}} \gamma^{\nu}_{\text{rad}} N^{\nu}_{\bm{Q}}(t) \label{eq:EoM} \\
 & + \Gamma^{\text{in}}_{\nu\bm{Q}}(t) - \Gamma^{\text{out}}_{\nu\bm{Q}} N^{\nu}_{\bm{Q}}(t) + \Gamma^{\text{form}}_{\nu\bm{Q}}(t) - \Gamma^{\text{diss}}_{\nu\bm{Q}} N^{\nu}_{\bm{Q}}(t). \nonumber
\end{align}
The first line in Eq.\,\eqref{eq:EoM} describes the interaction with light. Here, the first term accounts for the absorption of incident light expressed with the microscopic polarization $p^{\mu}(t)$, while the second term describes the loss of exciton occupation via radiative recombination\,\cite{brem2018exciton}. A more thorough description of exciton-light interaction can be found in the supplementary material.
The second line in Eq.\,\eqref{eq:EoM} accounts for exciton-phonon scattering within the second-order Born-Markov approximation\,\cite{brem2018exciton,kira2006many,kira2011semiconductor,haug2009quantum}.
The first two terms describe bound-to-bound transitions with the in- and out-scattering rates $\Gamma^{\text{in}}_{\nu\bm{Q}}(t)=\sum_{\mu\bm{Q}'}^{N_{\text{b}}} \Gamma^{\mu\nu}_{\bm{Q}'\bm{Q}} N^{\mu}_{\bm{Q}'}(t)$ and $\Gamma^{\text{out}}_{\nu\bm{Q}}=\sum_{\mu\bm{Q}'}^{N_{\text{b}}} \Gamma^{\nu\mu}_{\bm{Q}\bm{Q}'}$.
The appearing scattering matrix reads
$
 \Gamma^{\nu\mu}_{\bm{Q}\bm{Q}'} = \frac{2\pi}{\hbar} \sum_{j\pm}\left|G^{\nu\mu}_{j,\bm{q}}\right|^2 \eta^{j\pm}_{\bm{q}} \delta\left(E^{\mu}_{\bm{Q}'}-E^{\nu}_{\bm{Q}}\pm\hbar\Omega^j_{\bm{q}}\right).
$
Here, we have introduced the abbreviation $\eta^{j\pm}_{\bm{q}} = n^j_{\bm{q}}+\frac{1}{2}\pm\frac{1}{2}$ with $n^j_{\bm{q}} = \braket{b^{j\dagger}_{\bm{q}}b^{j\phantom{\dagger}}_{\bm{q}}}$ being the phonon number. Furthermore, $\bm{q}=\bm{Q}'-\bm{Q}$ is the momentum exchange, $\pm$ accounts for the emission ($+$) and absorption ($-$) of phonons, and $E^{\nu}_{\bm{Q}}$ and $\hbar\Omega^j_{\bm{q}}$ are the exciton and phonon energies, respectively.
While the first two terms in the second line of Eq.\,\eqref{eq:EoM} account for scattering within bound exciton states, the last two terms account for the formation/dissociation of excitons from/to free electron-hole pairs with the rates $\Gamma^{\text{form}}_{\nu\bm{Q}}=\sum_{\bm{pQ}'} \Gamma^{\bm{p}\nu}_{\bm{Q}'\bm{Q}} N^{\bm{p}}_{\bm{Q}'}$ and  $\Gamma^{\text{diss}}_{\nu\bm{Q}}=\sum_{\bm{pQ}'} \Gamma^{\nu\bm{p}}_{\bm{Q}\bm{Q}'}$ including a sum over all scattering possibilities.

We now make the assumption that free carriers are immediately extracted from the system after dissocitation due to a finite electric field resulting in a zero contribution from the exciton formation term. The electric field is assumed to be weak enough, so that field-induced dissociation is negligibly small.
Here, we focus on the microscopic origin of dissociation and its implications on the fundamental limits of real devices. Therefore, we assume a ballistic transport and a complete collection at the leads, so that the current is determined by the dissociation. The latter poses an upper fundamental limit for the maximum photocurrent one can obtain in a real device.  Furthermore, in addition to  the discussed terms appearing in Eq.\,\eqref{eq:EoM}, we also include a phenomenological decay $\dot{N}^{\nu}_{\bm{Q}}|_{\text{nr}} = -\tau^{-1}_{\text{nr}} N^{\nu}_{\bm{Q}}$ to account for non-radiative recombination, which can be significant in real samples due to defects\,\cite{zipfel2020exciton}. The decay time $\tau_{\text{nr}}$ has been experimentally and theoretically determined to range from few to hundreds of picoseconds\,\cite{zipfel2020exciton,massicotte2018dissociation,cadiz2018exciton,li2019defect}.

In this study, we discuss exciton dissociation in the four most studied semiconducting TMDs including WSe$_2$, WS$_2$, MoSe$_2$, and MoS$_2$. However, for now, we focus on hBN-encapsulated WSe$_2$ monolayers due to their well established excitonic landscape containing bright (KK) and momentum-dark ({K\textLambda} and KK') excitons\,\cite{zhang2015experimental,selig2018dark,tang2019long,rosati2020negative,rosati2020temporal}. We consider up to 7 bound s-states for each valley, assigning states with higher main quantum number to the continuum and disregarding states with non-zero angular momentum due to their lower exciton-phonon cross section\,\cite{brem2018exciton}.
Moreover, we exclude the influence of spin-dark states, since the timescale of spin-flip scattering processes is considerably slow comapared to the spin-conserving ones\,\cite{song2013transport,rosati2020temporal}. We use input parameters for the bandstructure and electron-phonon coupling from ab-initio studies\,\cite{kormanyos2015k,laturia2018dielectric,jin2014intrinsic,khatibi2018impact}, cf. the supplementary material.
\\

%%%%%%%%%%%%%%% DYNAMICS %%%%%%%%%%%%%%%%%%%%%%%%%%%%%%%%%%%%%%%%%%%%%%%%%%%%%%%%%%%%%%%%%%%%%%%%%
\noindent\textbf{Dissociation dynamics.}
Now, we resolve the exciton dissociation dynamics in  hBN-encapsulated WSe$_2$ monolayers. We solve  Eq.\,\eqref{eq:EoM} considering a continuous wave optical excitation resontant to the 1s exciton with a rise time of 1\,ps and a power density of 1 {\textmu}W/{\textmu}m$^2$. In order to account for a realistic sample with defects, we set the non-radiative decay time $\tau_{\text{nr}} = 100\,\text{ps}$, similar to experimentally reported values\,\cite{cadiz2018exciton}.
We define the dissociation current density $j_{\text{d}} = e_0 A^{-1}\sum_{\nu\bm{Q}}\Gamma^{\text{diss}}_{\nu\bm{Q}} N^{\nu}_{\bm{Q}}$ and disentangle the contributions from different channels, i.e. scattering from the state $\nu$ to the continuum of a given valley (cf. Fig.\,\ref{fig:Fig2}(a)). 
We identify the most important dissociation channel to be K\textLambda\textsubscript{2s}$\rightarrow$KK'\textsubscript{cont.} followed by K\textLambda\textsubscript{1s}$\rightarrow$KK'\textsubscript{cont.}. There is also a number of other dissociation channels with minor contributions that however sum up to account for a significant portion of the total current.
Furthermore, we find that the dissociation current increases on a timescale of 50--100\,ps, corresponding to $\tau^{-1}=\tau_{\text{diss}}^{-1}+\tau_{\text{nr}}^{-1}$. This is in agreement with the fact that the response time is dominated by the shortest decay time, cf. the supplementary material.
In the considered case, both dissociation and non-radiative decay show characteristic lifetimes of 100\,ps, and hence both dominate the response.
Although radiative recombination is very fast (below 1\,ps), radiative decay of excitons is rather slow (1-100 ns) because only excitons in the light cone ($\bm{Q} \approx 0$) can recombine\,\cite{selig2018dark} and hence radiative decay does not influence the response time.

The efficiency of the dissociation channel K\textLambda\textsubscript{bound}$\rightarrow$KK'\textsubscript{cont.} is a result of the strong coupling of excitons with M phonons\,\cite{jin2014intrinsic}, leading to a strong intervalley scattering from {K\textLambda} to KK' excitons.
This is manifested in high rates $\Gamma^{\text{diss}}_{\nu\bm{Q}}$ for  dissociation from K\textLambda\textsubscript{2s} compared to other exciton states with a similar energy (cf. Fig.\,\ref{fig:Fig2}(b)).
Moreover, the importance of K\textLambda\textsubscript{2s} over K\textLambda\textsubscript{1s} results from the interplay between occupation and dissociation rates: while lower states are largely occupied (cf. Fig.\,\ref{fig:Fig2}(e)), higher states exhibit faster dissociation rates due to their proximity to the continuum  (cf. Fig.\,\ref{fig:Fig2}(b)).
We also determine the main exciton pathway, starting from the optical excitation at KK\textsubscript{1s} and ending with the main dissociation channel K\textLambda\textsubscript{2s}$\rightarrow$KK'\textsubscript{cont.} identified above (cf. arrows in Fig.\,\ref{fig:Fig1}). For this purpose, we calculate the net scattering rates from $\nu$ to $\mu$ with $\Gamma_{\nu\mu} = A^{-1}\sum_{\bm{QQ}'}(\Gamma^{\nu\mu}_{\bm{QQ}'}N^{\nu}_{\bm{Q}}-\Gamma^{\mu\nu}_{\bm{Q}'\bm{Q}}N^{\mu}_{\bm{Q}'})$, as illustrated in Fig.\,\ref{fig:Fig2}(c). We show that most excitons follow the following path: after optical excitation of KK\textsubscript{1s} excitons, they scatter to K\textLambda\textsubscript{1s} states  followed by K\textLambda\textsubscript{1s}$\rightarrow$KK'\textsubscript{2s} and KK'\textsubscript{2s}$\rightarrow$K\textLambda\textsubscript{2s} transitions, and finally they dissociate through K\textLambda\textsubscript{2s}$\rightarrow$KK'\textsubscript{cont.}. Thus, not only is the dissociation dominated by scattering with M phonons, but also the transitions within bound states, which occur mostly between  KK' and {K\textLambda} states.

\begin{figure}[t!]
 \centering
 \includegraphics[width=\linewidth]{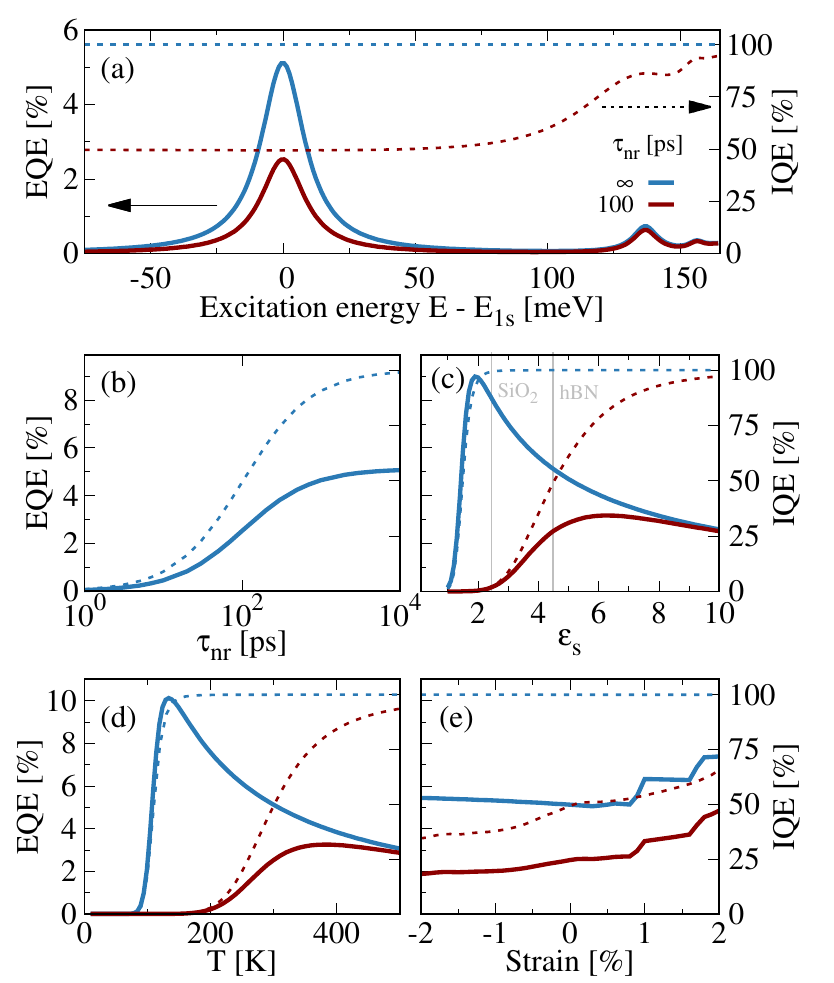}
 \caption{\textbf{Quantum efficiency}. External (solid lines, left axis) and internal (dashed lines, right axis) quantum efficiencies as a function of (a) excitation energy, (b) non-radiative recombination time, (c) dielectric constant of the substrate, (d) temperature, and (e) strain assuming a non-radiative decay time $\tau_{\text{nr}} \to \infty$ (blue) and $\tau_{\text{nr}} = 100\,\text{ps}$ (red).
}
 \label{fig:Fig3}
\end{figure}

Exploiting the microscopic character of our theoretical approach, we can resolve the dynamics of excitons in energy, momentum and time. At short times during the start of the optical excitation ($t = 0.1\,\text{ps}$, Fig.\,\ref{fig:Fig2}(d)), a small occupation has already been generated at the 1s level of all valleys as a result of ultrafast polarization-to-population transfer and inter-valley scattering, with sharp features appearing due to scattering with optical {\textGamma} phonons and intervalley {\textLambda} and M phonons\,\cite{selig2018dark,brem2018exciton}. Higher KK states (2s, 3s) show minor occupation due to off-resonant excitation. At later times, when the stationary state has nearly been reached ($t = 200\,\text{ps}$, Fig.\,\ref{fig:Fig2}(e)), the occupation shows a more thermalized distribution, with larger occupation in higher exciton states compared to earlier times, and a strong occupation in the light cone at $Q=0$  due to the continuous optical excitation. The equilibration between bound and unbound states described here is opposite to the relaxation cascade\,\cite{brem2018exciton} but occurs on the same timescale of a few picoseconds.
 Finally, we compare the exciton occupation at $t = 200\,\text{ps}$ with the case where dissociation has not been considered ($N^{\text{nd}}_{\bm{Q}}$). In order to be able to compare the occupation in these two cases, we normalize them by the total density $n$, i.e. $\tilde{N}_{\bm{Q}} = N_{\bm{Q}}/n$. We compute the relative difference $(\tilde{N}_{\bm{Q}}-\tilde{N}^{\text{nd}}_{\bm{Q}})/\tilde{N}^{\text{nd}}_{\bm{Q}}$ and find that higher states are clearly less occupied when the effect of dissociation is taken into account (cf. Fig\,\ref{fig:Fig2}(f)). The lower occupation of higher states is a direct consequence of their efficient dissociation (cf. Fig\,\ref{fig:Fig2}(b)).\\

%%%%%%%%%%%%%%% EFFICIENCY AND RESPONSE TIME %%%%%%%%%%%%%%%%%%%%%%%%%%%%%%%%%%%%%%%%%%%%%%%%%%%%%%%%%%%%%%
\noindent\textbf{Quantum efficiency and response time.}
The rise time of the dissociation current is limited by the fastest decay mechanism. An efficient dissociation can hence pose a fundamental limit on how fast the response time can be in a real device. Moreover, the magnitude of the dissociation current represents as well a fundamental limit for the internal (IQE) and external (EQE) quantum efficiencies. In real devices, the response time will be longer and the quantum efficiency will be lower because of imperfect charge transport and collection at the leads.
In this work, we show the optimal efficiency and response time that can be reached in 2D material based photodetectors.

\begin{figure}[t!]
 \centering
 \includegraphics[width=\linewidth]{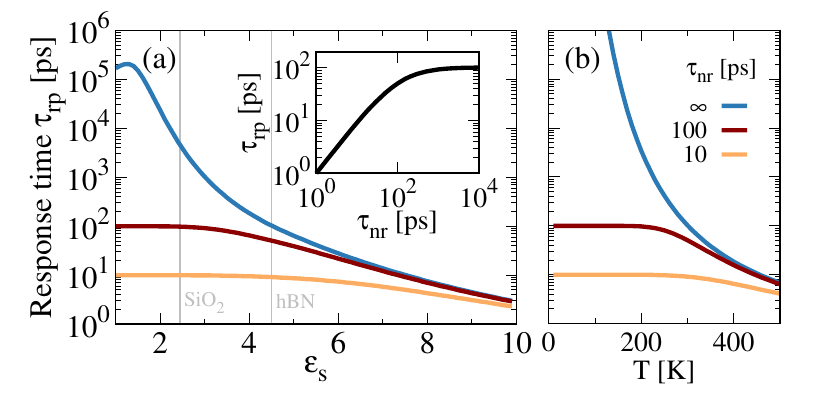}
 \caption{\textbf{Response time}. Response time as a function of the (a) dielectric constant of substrate and (b) temperature for $\tau_{\text{nr}} \to \infty$ (blue), $\tau_{\text{nr}} = 100\,\text{ps}$ (red) and $\tau_{\text{nr}} = 10\,\text{ps}$ (orange). The inset in (a) shows the response time as a function of non-radiative decay time $\tau_{\text{nr}}$.}
 \label{fig:Fig4}
\end{figure}

In the last section, we have resolved and understood the microscopic channels responsible for exciton dissociation. Here, we investigate their impact on key quantities for a photodetector: EQE, IQE and the response time.
EQE and IQE describe the efficiency of dissociation with respect to the rate of incident photons $\Phi=\frac{I}{\hbar\omega}$ and of optically generated excitons $G$, respectively, and are directly related through the optical absorption $\alpha(\omega)$, i.e. $\text{EQE} = \alpha(\omega) \text{IQE}$.
Moreover, the response time $\tau_{\text{rp}}$ is determined by the dominant decay time, which can be non-radiative decay ($\tau_{\text{nr}}$), dissociation ($\tau_{\text{d}}$), or radiative decay ($\tau_{\text{r}}$), cf. supplementary material.
Thus, we can write
\begin{equation}
 \label{eq:QE}
 \text{IQE}=\dot{n}_{\text{d}}/G,\quad \text{EQE}=\dot{n}_{\text{d}} \hbar\omega / I,\quad \tau_{\text{rp}}^{-1} = \sum_i \tau_i^{-1},
\end{equation}
where $\omega$ and $I$ are the frequency and power density of the incident light, respectively, and $i=\{\text{nr},\text{d},\text{r}\}$. From our microscopic model, we obtain the exciton generation rate $G=A^{-1} \sum_{\mu\nu\bm{Q}} \Gamma^{\mu\nu}_{\bm{0Q}} |p^{\mu}(t)|^2=\frac{I}{\hbar\omega}\alpha(\omega)$ and the dissociation rate $\dot{n}_{\text{d}}=A^{-1}\sum_{\nu\bm{Q}}\Gamma^{\text{diss}}_{\nu\bm{Q}} N^{\nu}_{\bm{Q}}$.

In the following, we reveal how quantum efficiency and response time can be controlled and optimized through externally accessible knobs, such as excitation energy, non-radiative decay time, substrate screening, temperature and strain.
We model the effect of defects by investigating two cases of non-radiative decay time: $\tau_{\text{nr}} \to \infty$ and $\tau_{\text{nr}} = 100\,\text{ps}$, accounting for a pristine defect-free sample and a more realistic sample with defects\,\cite{cadiz2018exciton}, respectively. Unless otherwise stated, the excitation energy is centered at the 1s exciton resonance, the dielectric environment corresponds to hBN encapuslation, the temperature is 300 K and the sample is unstrained.
We solve Eq.\,\eqref{eq:EoM} in the stationary state, setting $\dot{N}^{\nu}_{\bm{Q}}=0$, and use the calculated exciton occupations to determine the EQE, IQE, and the response time $\tau_{\text{rp}}$ in Eq.\,\eqref{eq:QE}.

In a pristine defect-free sample, dissociation is the fastest decay mechanism and hence all optically generated excitons dissociate, yielding an IQE of 100\,\%. The EQE follows the absorption spectrum for varying excitation energies (cf. blue solid line in Fig.\,\ref{fig:Fig3}(a)), with maxima at the exciton resonance energies (1s, 2s, 3s). When including a realistic non-radiative recombination time of 100\,ps, the IQE drops down to 50\,\%, and the EQE decreases by the same factor (cf. red lines in Fig.\,\ref{fig:Fig3}(a)). The IQE is constant for a wide range of energies where the 1s exciton is excited. As the excitation energy approaches the 2s resonance, excitons in this state are excited as well. Since dissociation from higher states is more efficient (cf. Fig.\,\ref{fig:Fig2}(b)), the IQE increases with excitation energy, though showing again a plateau around  2s and 3s resonances (cf. the red dashed line in Fig.\,\ref{fig:Fig3}(a)). 
The most optimal performance is thus obtained by exciting at excitonic resonances, where the optical absorption is maximized, with a trade-off between lower absorption and faster dissociation at higher states.

The effect of defects can be studied by varying the non-radiative decay time $\tau_{\text{nr}}$ (cf. Fig.\,\ref{fig:Fig3}(b)). When non-radiative recombination is much faster than dissociation, most excitons are lost via this decay channel and the IQE is 0\,\%. As $\tau_{\text{nr}}$ progressively approaches the dissociation time $\tau_{\text{diss}}\approx 100$\,ps, dissociation starts to take over and the IQE increases. At even longer $\tau_{\text{nr}}>10^3$\,ps dissociation dominates yielding an IQE of 100\,\%. Nevertheless, there is a trade-off between efficiency and response time. For short $\tau_{\text{nr}}$, the response time will be dominated by this value and can be tuned, while the quantum efficiency becomes low (cf. inset in Fig.\,\ref{fig:Fig4}(a)). Remarkably, similar response times on the order of 100\,ps have been also reported in experimental studies for the same system at low electric fields in Ref.\,\onlinecite{massicotte2018dissociation}, where $\tau_{\text{nr}} \sim 300$\,ps.
It was shown that the measured response times deviate from the field-assisted tunneling model at low electric fields, showing saturation at values around 100\,ps. This deviation is consistent with our phonon-induced dissociation model and indicates that exciton-phonon scattering indeed limits the photoresponse at weak fields. Our theoretical predictions thus have important implications on the limitations of real TMD-based optoelectronic devices.

Another experimentally accessible knob is dielectric engineering via substrate screening that influences the material's charateristics in many different ways. First, stronger screening results in lower exciton binding energies, i.e. the excitonic levels are closer to each other and to the continuum, leading to a more efficient exciton dissociation. This is manifested in a significant increase of IQE and EQE (cf. Fig.\,\ref{fig:Fig3}(c)) and a decrease of the response time (Fig.\,\ref{fig:Fig4}(a)). On the other hand, screening also affects the absorption and thus the EQE. The optical absorption is directly proportional to the inverse dielectric function, $\alpha \propto \varepsilon^{-1}_{\text{s}}$. Moreover, the oscillator strength is proportional to the exciton probability $|\phi^{\nu}(\bm{r}=0)|^2$, which also decreases for stronger screening due to more delocalized excitonic wavefunctions. Both facts contribute to a reduction of the absorption, resulting in the decrease of the EQE at large dielectric screening constants, cf. Fig.\,\ref{fig:Fig3}(c).
The reduction of the oscillator strength also weakens radiative recombination, causing the response time to slightly increase for low $\varepsilon_{\text{s}}$ in a pristine sample (cf. blue line in Fig.\,\ref{fig:Fig4}(a)). For weak screening, dissociation is so inefficient that the main decay channel is radiative recombination in a defect-free sample.
Including a finite $\tau_{\text{nr}}$ of 100\,ps results in an overall reduction of the efficiency (cf. red lines in Fig.\,\ref{fig:Fig3}(c)). However, while dissociation is very inefficient on a SiO$_2$ substrate ($\tau_{\text{d}} \sim 4\ \text{ns}$), the IQE increases to approximately 50\%  on an hBN-encapsulated sample and the dissociation can compete with non-radiative recombination. The most optimal quantum efficiency is thus obtained for substrates with a stronger screening (such as hBN) giving rise to a more efficient exciton dissociation.

A similar behaviour is observed when exploring temperature as a potential knob to tune the performance of a photodetector. Here, the boost in the IQE (Fig.\,\ref{fig:Fig3}(d)) and the drop in response time (Fig.\,\ref{fig:Fig4}(b)) are caused by an increase in the phonon number $n^j_{\bm{q}}$ with temperature. As more phonons are available, exciton-phonon scattering becomes stronger, populating higher excitonic states and increasing the  dissociation efficiency. Note that at very low temperatures there is a 0\,\% plateau in the IQE and EQE at which radiative recombination is more efficient than dissociation. While the IQE displays a monotonic increase with temperature, the EQE reaches a maximum and then decreases due to lower absorption at higher temperatures. The increase in the phonon-induced linewidth of excitonic resonances\,\cite{selig2016excitonic} results in a broader absorption. Since the oscillator strength is conserved, the absorption peak decreases causing a clear reduction of the EQE at the exciton resonance (cf. solid lines in Fig.\,\ref{fig:Fig3}(d)).
The inclusion of a finite $\tau_{\text{nr}}$ shifts the IQE plateau to higher temperatures, at which exciton dissociation can compete with non-radiative decay, giving rise to an overall decrease of the EQE at low temperatures (cf. red lines in Fig.\,\ref{fig:Fig3}(d)).
Overall, the optimal quantum efficiency is found for temperatures slightly above room temperature for a realistic finite non-radiative decay.

\begin{figure}[t!]
 \centering
 \includegraphics[width=\linewidth]{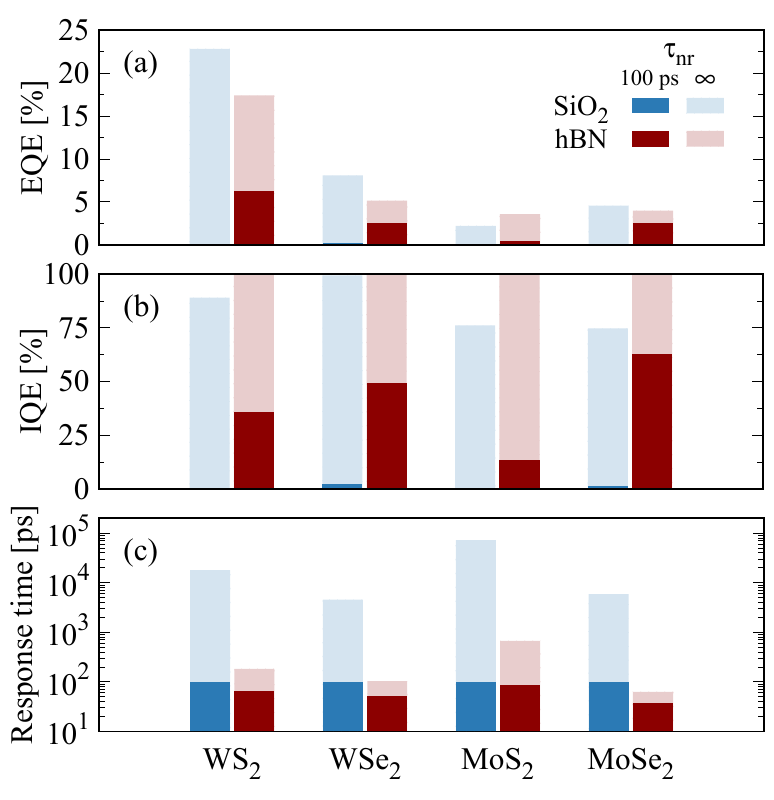}
 \caption{\textbf{Comparison of TMD materials}. (a) External quantum efficiency (EQE), (b) internal quantum efficiency (IQE) and (c) response time for four TMDs on SiO\textsubscript{2} substrate (blue) and in hBN encapsulation (red) studied for two non-radiative decay times $\tau_{\text{nr}}=100\,\text{ps}$ (dark color) and $\tau_{\text{nr}}\to\infty$ (light color).}
 \label{fig:Fig5}
\end{figure}

Finally, we study the effect of strain, which is known to shift the electronic K and {\textLambda} valleys in opposite directions with respect to the valence band\,\cite{khatibi2018impact}. These shifts have a direct impact on the linewidth and thus on the EQE of a pristine sample. For increasing tensile strain at approximately  +1\,\%, the K\textLambda\textsubscript{1s} state approaches the KK\textsubscript{1s} state, suppressing the scattering channel KK\textsubscript{1s}$\rightarrow$K\textLambda\textsubscript{1s} induced by phonon emission. Similarly, further increasing strain towards +2\,\% results in a shift of K\textLambda\textsubscript{1s} above KK\textsubscript{1s} and the inhibition of the intervalley scattering induced by phonon absorption. As a result, the exciton linewidth decreases and the absorption peak increases with tensile strain resulting in a step-wise increase of the EQE whenever a scattering channel is blocked (cf. solid blue line in Fig.\,\ref{fig:Fig4}(e)). The slight increase in the EQE at negative (compressive) strain can be traced back to a larger spectral distance between the KK and {K\textLambda} excitons, which shifts the final state $\ket{\mu,\bm{q}}$ to larger momenta, resulting in a smaller wavefunction overlap $\mathcal{F}^{\nu\mu}_{\bm{q}}$ for the intervalley scattering.
In order to understand the effect of strain in a  more realistic sample with defects ($\tau_{\text{nr}}= 100\,\text{ps}$), we study the changes in the IQE, which reflects the efficiency of dissociation with respect to the non-radiative decay. For tensile strain, the {K\textLambda} valley is lifted up so that the K\textLambda\textsubscript{1s} state lies closer to the continuum of the KK' valley. This optimal situation favours the dissociation of excitons from K\textLambda\textsubscript{1s} to KK'\textsubscript{cont.}, yielding an enhancement of the IQE (cf. dashed red line in Fig.\,\ref{fig:Fig4}(e)). In contrast, when compressive strain is applied, the {K\textLambda} valley shifts down. As a result,  K\textLambda\textsubscript{1s} has by far the largest occupation. Since this state moves further away from KK'\textsubscript{cont.} as compressive strain increases, the dissociation becomes less efficient and the IQE decreases.

Considering realistic values for non-radiative decay time, the quantum efficiency for exciton dissociation is the highest and the response time the fastest at the following conditions: temperatures slightly above 300K,  substrates with strong dielectric screeening, excitation energies resonant to exciton states, and tensile strain values around 1\%. For short non-radiative decay times, there is a trade-off between a fast response time and a low quantum efficiency.
\\

%%%%%%%%%%%%%%% TMD COMPARISON %%%%%%%%%%%%%%%%%%%%%%%%%%%%%%%%%%%%%%%%%%%%%%%%%%%%%%%%%%%%%%%%%%%%%%%%%
\noindent\textbf{Comparison of TMD materials.}
So far, we have focused on exciton dissociation in a WSe$_2$ monolayer encapsulated in hBN. Now, we compare EQE, IQE and the response time for the four most studied semiconducting TMDs including WSe$_2$, WS$_2$, MoSe$_2$, and MoS$_2$ placed on the two most common dielectric environments (SiO$_2$ substrate  and hBN encapsulation), cf. Fig.\,\ref{fig:Fig5}. 
The highest EQE in a defect-free monolayer (i.e. $\tau_{\text{nr}}\to\infty$, cf. light colors in Fig.\,\ref{fig:Fig5}) on a SiO$_2$ substrate is found for WS$_2$ (23\,\%) due to its large optical absorption\,\cite{li2014measurement}, followed by WSe$_2$ (8\,\%), MoSe$_2$ (5\,\%), and MoS$_2$ (2\,\%). Since here the dissociation is faster than the radiative decay, the IQE is close to 100\,\% in the absence of non-radiative recombination.
Despite the relatively large EQE and IQE values, exciton dissociation in TMDs on a SiO$_2$ substrate is very inefficient as a result of the large binding energies and leads to very long response times ranging from few to tens of nanoseconds. Due to such inefficiency, including a non-radiative decay time of 100\,ps completely suppresses dissociation, resulting in quantum efficiencies much below 1\,\%, which are barely visible in Figs.\,\ref{fig:Fig5}(a)-(b).
Under hBN encapsulation, the excitonic oscillator strength decreases resulting in a weaker optical absorption and hence in lower EQE values. In MoS$_2$, however, this effect is countered by the suppression of the intervalley KK$\rightarrow${\textGamma}K dephasing channel via phonon emission. This is due to a lower energetic separation between the two valleys, resulting in a much smaller linewidth and hence a larger absorption and EQE.
In hBN encapsulated samples, exciton dissociation is much more efficient because of the smaller binding energies and yields response times on the order of 100\,ps. Including a realistic non-radiative decay of 100\,ps causes the EQE and IQE to decrease to moderate values around $2-6\,\%$ and $30-60\,\%$ respectively, including a modest improvement of the response time to values below 100\,ps. 

In order to understand why dissociation is more efficient in some TMDs, we need to consider their complex excitonic landscape.
The long response time in pristine MoS$_2$ on any substrate compared to the other TMDs is a consequence of the slow exciton dissociation in this material (cf. Fig.\,\ref{fig:Fig5}(c)).
This can be traced back to the fact that the largest occupation is found in   \textGamma{K} and \textGamma{K'} excitons exhibiting a large effective mass and hence large exciton binding energies. Thus, it is difficult to scatter into higher exciton states, which are energetically far away. 
Note however that the excitonic landscape and in particular the question which states is the lowest is still under debate in the case of MoS$_2$ \,\cite{deilmann2019finite,mai2014many,steinhoff2015efficient,feierabend2020brightening}. 

The TMD material with the fastest dissociation on a SiO$_2$ substrate is WSe$_2$ (cf. light blue boxes in Fig.\,\ref{fig:Fig5}(c)) due to its optimal band structure favouring the dissociation channel K\textLambda\textsubscript{2s}$\rightarrow$KK'\textsubscript{cont.}. Under hBN encapsulation, however, it is MoSe$_2$ that shows the fastest dissociation due to the proximity of  K\textLambda\textsubscript{1s} excitons to the KK' continuum.
In both dielectric environments, the dissociation in WS$_2$ is not as fast as in WSe$_2$ and MoSe$_2$. While the band structure of WS$_2$ is similar to the one of WSe$_2$, the main state for dissociation K\textLambda\textsubscript{2s} is too high in energy exhibiting a very low  occupation and thus giving rise to a longer dissociation time.
Note that the dielectric environment and in particular hBN encapsulation will also modify the transport properties affecting the performance of a real device. Our calculations provide an upper fundamental limit of the performance limited by dissociation.
\\

%%%%%%%%%%%%%%% CONCLUSIONS %%%%%%%%%%%%%%%%%%%%%%%%%%%%%%%%%%%%%%%%%%%%%%%%%%%%%%%%%%%%%%%%%%%%%%%%%
\noindent\textbf{Conclusions.}
Using a microscopic approach, we have resolved the complex many-particle processes behind phonon-assisted exciton dissociation  in TMD monolayers exhibiting a multi-valley exciton landscape. Focusing on WSe$_2$, we have discerned the distinct transition channels dominating exciton dissociation, which involve phonon-driven intervalley scattering from {K\textLambda} to KK' excitons. Due to the trade-off between lower occupation and larger dissociation rates at higher exciton states, most of the dissociation arises from the 2s exciton of the {K\textLambda} valley. 
Furthermore, we have determined fundamental limits for the efficiency and response time of TMD-based optoelectronic devices.
In particular, we find a dissociation-limited response time of 100\,ps for hBN-encapsulated WSe$_2$ monolayer, which is similar to experimental findings.
Furthermore, we have studied the tunability of key quantities with externally accessible knobs such as excitation energy, dielectric engineering, temperature and strain on a microscopic footing. We find a trade-off between faster response time and lower EQE for increasing substrate screening, temperature, non-radiative decay rate, and excitation energy. 
\\[10pt]

%%%%%%%%%%%%%%% ACKNOWLEDGEMENTS %%%%%%%%%%%%%%%%%%%%%%%%%%%%%%%%%%%%%%%%%%%%%%%%%%%%%%%%%%%%%%%%%%%%%%%%%
\noindent\textbf{Acknowledgments}\\
This project has received funding from the Swedish Research Council (VR, project number 2018-00734), the  European Union’s Horizon 2020 research and innovation programme under grant agreement no. 881603 (Graphene Flagship), and Chalmer's Excellence Initiative Nano under its excellence PhD programme.
The computations were enabled by resources provided by the Swedish National Infrastructure for Computing (SNIC) at C3SE partially funded by the Swedish Research Council through grant agreement no. 2016-07213. 
We thank Thomas Müller (TU Wien) and Roberto Rosati (Chalmers) for fruitful discussions.
%%%%%%%%%%%%%%%%%%%%%%%%%%%%%%%%%%%%%%%%%%%%%%%%%%%%%%%%%%%%%%%%%%%%%%%%%%%%%%%%%%%%%%%%%%%%%%%%%%%%%%%%%%

%\bibliographystyle{achemso}
%\bibliography{ref}

%% COPIED FROM .bbl FILE:


\iffalse
%% TOC
\begin{minipage}{\linewidth}
 \centering
 \includegraphics[scale=1]{./TOC_figure.pdf}
 Graphical TOC.
\end{minipage}
\fi

\end{document}